\documentstyle[11pt,aaspp4]{article}  

\begin{document}

\title{Time-Resolved Photometry of Kuiper Belt Objects: \\
Rotations, Shapes and Phase Functions}  
\author{Scott S. Sheppard and David C. Jewitt}    
\affil{Institute for Astronomy, University of Hawaii, \\
2680 Woodlawn Drive, Honolulu, HI 96822 \\ sheppard@ifa.hawaii.edu, jewitt@ifa.hawaii.edu}

\begin{abstract}  

We present a systematic investigation of the rotational lightcurves of
trans-Neptunian objects based on extensive optical data from Mauna
Kea.  Four of 13 objects (corresponding to 31\%) in our sample
((33128) 1998 BU$_{48}$, 2000 GN$_{171}$, (20000) Varuna and 1999
KR$_{16}$) were found to exhibit lightcurves with peak-to-peak range
$\geq 0.15$ magnitude.  In a larger sample obtained by combining our
data with reliably determined lightcurves from the literature, 7 of 22
objects (32\%) display significant ($\geq 0.15$ magnitude range)
lightcurves.  About $23 \%$ of the sampled objects have lightcurve
ranges $\geq 0.4$ magnitudes.  Curiously, the objects are very large
($\gtrsim$ 250 km diameter, assuming an albedo of 0.04) and, in the
absence of rotation, should be near spherical due to self compression.
We propose that the large amplitude, short period objects are
rotationally distorted, low density rubble piles.  Statistically, the
trans-Neptunian objects are less spherical than their main-belt
asteroid counterparts, indicating a higher specific angular momentum
perhaps resulting from the formation epoch.  In addition to the
rotational lightcurves, we measured phase darkening for $7$ Kuiper
Belt objects in the $0$ to $2$ degree phase angle range.  Unlike
Pluto, the measured values show steep slopes and moderate opposition
surge indicating backscatter from low albedo porous surface materials.

\end{abstract}

\keywords{Kuiper Belt, Oort Cloud - minor planets, solar system: general}

\newpage

\section{Introduction}

More than 500 Trans-Neptunian Objects (TNOs) have been discovered in the
decade since the discovery of 1992 QB$_{1}$ (Jewitt \& Luu 1993).
These objects comprise the Kuiper Belt (also known as the
Edgeworth-Kuiper Belt) which is thought to contain about 70,000
objects with radii greater than 50 km (Jewitt, Luu and Chen 1996).
The Kuiper Belt is thought to be a relic from the original
protoplanetary disk, albeit one that has been dynamically disturbed and
collisionally processed in ways that are not yet fully understood.

The Kuiper Belt is the most likely source of the Jupiter-family comets
(Fernandez 1980, Duncan, Quinn and Tremaine 1988).  It is by far the
largest long-lived reservoir of small bodies in the planetary region,
outnumbering the main-belt asteroids and Jovian Trojans by a factor of
$\sim$ 300.  The Kuiper Belt Objects (KBOs) are further thought to be
chemically primitive, containing trapped volatiles and having
experienced relatively little thermal evolution since formation.  Thus
we may be able to probe some aspects of the early history of the local
solar nebula by studying the Kuiper Belt and related objects.

The determination of the physical characteristics of the KBOs has
proceeded very slowly.  This is because even the brightest known KBOs
(other than Pluto and Charon) reach only apparent red magnitude
$m_{R}\sim 19.5$ and thus are challenging with current spectroscopic
technology.  The surfaces of KBOs may have been altered over their
lifetimes by collisions, cometary activity, and irradiation.  The
largest KBOs might even be partially differentiated from radiogenic
heating.  This could lead to the spinning up of objects to conserve
angular momentum.  Colors of the KBOs have been found to be diverse,
ranging from neutral to very red (V-R$ \sim 0.3$ to V-R$\sim 0.8$)
(Luu \& Jewitt 1996; Green et al. 1997; Tegler \& Romanishin 2000;
Jewitt \& Luu 2001).  While spectra of KBOs are mostly featureless,
some show weak 2$\mu$m water ice absorptions (Brown, Cruikshank, \&
Pendleton 1999; Jewitt \& Luu 2001).  Most KBOs are too distant
($\gtrsim 30$ AU) and small to resolve with current technology.  They
are also very cold objects ($\sim 50$K) which emit most of their
thermal radiation in the inaccessible far infrared wavelengths,
requiring observations from above the Earth's atmosphere.  Thus the
most feasible way to determine KBOs shapes and surface features is
through their photometric light variations.

The rotations and shapes of the KBOs may be a function of their size.
Small KBOs (diameters $D <$ 100 km) are thought to be
collisionally produced (Farinella and Davis 1996).  These objects
retain no memory of the primordial angular momentum of their parent
bodies.  Instead, their spins are presumably set by the partitioning
of kinetic energy delivered by the projectile responsible for
break-up.  Larger objects may be structurally damaged bodies held
together by gravity (rubble piles).  The spins of these objects should
be much less influenced by recent impacts.  A similar situation
prevails in the main asteroid belt, where collisional modification of
the rotations and shapes of the smaller objects is observationally
well established (Catullo et al. 1984).  The large objects in both the
main-belt and the Kuiper Belt may provide a record of the primordial
distribution of angular momenta imbued by the growth process. A key
attribute of the Kuiper Belt is that the population is very large
compared to the main asteroid belt, allowing access to a substantial
sample of objects that are too large to have been influenced by recent
collisions.

We here use voluminous time resolved photometric observations to
determine the rotational lightcurves, colors, and phase functions of
KBOs.  As our sample, we select the intrinsically brightest
(presumably largest) KBOs. Specifically, we observed KBOs having
absolute magnitude $H_R \leq 7.5$, corresponding to $D \geq$ 200 km if
a red geometric albedo of $p_R$ = 0.04 is assumed.  We use most of the
known KBOs with $H_R \leq 6.0$ which corresponds to $D \geq$ 375 km in
our analysis.  The objects observed were all bright in order to
guarantee high signal-to-noise ratios in short exposures to adequately
sample the KBO lightcurves.

\section{Observations}

The University of Hawaii 2.2 m diameter telescope atop Mauna Kea in
Hawaii was used with a $2048 \times 2048$ pixel Tektronix CCD ($24$
$\micron$ pixels) and a $0.\arcsec 219$ pixel$^{-1}$ scale at the f/10
Cassegrain focus.  An antireflection coating provides very high average
quantum efficiency (0.90) in the R-band.  The field-of-view was
$7\arcmin .5 \times 7\arcmin .5$.  Exposures were taken using BVRI
filters based on the Johnson-Kron-Cousins system, while the telescope
was autoguided on bright nearby stars.  The seeing ranged from
$0.\arcsec 6$ to $1.\arcsec 5$ during the many nights of observation
throughout 1999, 2000, and 2001.  Objects moved relative to the fixed
stars at a maximum of $4\arcsec$ hr$^{-1}$ corresponding to trail
lengths $\leq$ $0.\arcsec 45 $ in the longest (400 sec) exposures.
Even for the fastest moving objects in the longest exposures the
trailing motion is small compared to the seeing and so can be neglected 
as a source of error in the photometry.

The images were bias subtracted and then flat-fielded using the median
of a set of dithered images of the twilight sky.  Landolt (1992)
standard stars were used for the absolute photometric calibration.
Photometry of faint objects, such as the KBOs, must be done very
carefully to achieve accurate results.  To optimize the
signal-to-noise ratio we performed aperture correction photometry by
using a small aperture on the KBOs ($0.\arcsec 65$ to $0.\arcsec 88$
in radius) and both the same small aperture and a large aperture
($2.\arcsec 40$ to $3. \arcsec 29$ in radius) on (four or more) nearby
bright field stars.  We corrected the magnitude within the small
aperture used for the KBOs by determining the correction from the
small to the large aperture using the field stars (c.f. Tegler and
Romanishin 2000; Jewitt \& Luu 2001).  Since the KBOs moved slowly we
were able to use the same field stars from night to night within each
observing run.  Thus relative photometric calibration from night to
night was very constant.  The few observations that were taken in
mildly non-photometric conditions were calibrated to observations of
the same field stars on the photometric nights.  The observational
circumstances, geometry, and orbital characteristics of the 13
observed KBOs are shown in Tables~1 and 2 respectively.

\section{Lightcurve Results}

The photometric results for the 13 KBOs are listed in Table~3, where
the columns include the start time of each integration, the
corresponding Julian date, and the magnitude.  No correction for light
travel time has been made.  Results of the lightcurve analysis for all
the KBOs observed are summarized in Table~4 while the mean colors can
be found in Table~5.  We first discuss the lightcurves of (20000)
Varuna, 2000 GN$_{171}$, (33128) 1998 BU$_{48}$, and 1999 KR$_{16}$
and give some details about the null results below.

We employed the phase dispersion minimization (PDM) method
(Stellingwerf 1978) to search for periodicity in the data.  In PDM, the
metric is the so-called $\Theta$ parameter, which is essentially the
variance of the unphased data divided by the variance of the data when
phased by a given period.  The best fit period should have a very small
dispersion compared to the unphased data and thus $\Theta <<$ 1
indicates that a good fit has been found.

\subsection{(20000) Varuna}
Varuna shows a large, periodic photometric variation (Farnham 2001).
We measured a range $\Delta m_R$ = 0.42 $\pm$ 0.02 mag. and best-fit,
two-peaked lightcurve period $P$ = 6.3436 $\pm$ 0.0002 hrs (about
twice the period reported by Farnham), with no evidence for a
rotational modulation in the $B-V$, $V-R$ or $R-I$ color indices.
These results, and their interpretation in terms of a rotating,
elongated rubble pile of low bulk density, are described in detail in
Jewitt and Sheppard (2002).

\subsection{2000 GN$_{171}$}

PDM analysis shows that 2000 GN$_{171}$ has strong PDM minima near
periods $P$ = 4.17 hours and $P$ = 8.33 hours, with weaker 24 hour
alias periods flanking each of these (Figure~\ref{fig:gnpdm}).  We
phased the data to all the peaks with $\Theta < 0.4$ and found only
the 4.17 and 8.33 hour periods to be consistent with all the data.
The $P$ = 4.17 hour period gives a lightcurve with a single maximum
per period while the $P$ = 8.33 hour lightcurve has two maxima per
period as expected for rotational modulation caused by an aspherical
shape.  Through visual inspection of the phased lightcurves we find
that the phase plot for $P$ = 4.17 hour
(Figure~\ref{fig:lightgnsingler}) is more scattered than that for the
longer period of $P$ = 8.33 hour (Figure~\ref{fig:lightgnr}).  This is
because the double-peaked phase plot shows a significant asymmetry of
$\Delta \sim 0.08$ magnitudes between the two upper and lower peaks.
A closer view of the PDM plot in Figure~\ref{fig:gnpdmsmall} around
the double-peaked period allows us to obtain a rotation period of
$P_{rot} = 8.329 \pm 0.005$ hours with a peak-to-peak variation of
$\Delta m = 0.61 \pm 0.03$ magnitudes.  We believe that the
photometric variations in 2000 GN$_{171}$ are due to its elongated
shape rather than to albedo variations on its surface.

Broadband BVRI colors of 2000 GN$_{171}$ show no variation throughout
its rotation within the photometric uncertainties of a few \%
(Figures~\ref{fig:lightgnbvri} and~\ref{fig:gncolors} and Table~6).
This again suggests that the lightcurve is mostly caused by an
elongated object with a nearly uniform surface.  The colors
$B-V=0.92\pm 0.04$, $V-R=0.63\pm 0.03$, and $R-I=0.56\pm 0.03$
(Table~5 and Table~6) show that 2000 GN$_{171}$ is red but
unremarkably so as a KBO (Jewitt and Luu 2001).

\subsection{(33128) 1998 BU$_{48}$}
The KBO 1998 BU$_{48}$ showed substantial variability ($>0.4$
magnitude with period $>4.0$ hour) in R-band observations from 2
nights in 2001 February and April. However, a convincing lightcurve
could not be found from just these 2 nights separated by 2 months.
Additional observations were obtained in the period 2001 November
$14-19$.  One minimum and one maximum in brightness within a single
night was observed and put the full single-peaked lightcurve between
about 4 and 6 hours.  Through PDM analysis, 1998 BU$_{48}$ was found
to have a peak-to-peak variation of $\Delta m = 0.68\pm 0.04$
magnitudes with possible single-peaked periods near 4.1, 4.9, and 6.3
hours which are 24 hour aliases of each other
(Figure~\ref{fig:bupdm}).  By examining the phased data using these
three possible periods we find that the single-peaked periods of
$4.9\pm 0.1$ and $6.3\pm 0.1$ hours are both plausible
(Figure~\ref{fig:buphase}).  The colors, $B-V=0.77\pm 0.05$,
$V-R=0.68\pm 0.04$, and $R-I=0.50\pm 0.04$ (Table~5) show no sign of
variation throughout the lightcurve, within the measurement
uncertainties (Table~7 and Figure~\ref{fig:buphase}).

\subsection{1999 KR$_{16}$}

This object was observed on four different observing runs during the
course of 2000 and 2001.  The data from 2001 are more numerous and of
better quality than the data from 2000.  We observed one brightness
minimum and one maximum within a single night of data and from this
estimated that the full single-peaked lightcurve should be near 6
hours.  In a PDM plot constructed using only the inferior data from
2000 we found single-peaked minima at 4.66 and 5.82 hours.  Phased
lightcurves at these periods are acceptable for the year 2000 data,
but the 4.66 hour period is inconsistent with the data from 2001.  In
the PDM plot using the R-band data from February, April, and May 2001
the best fit single-peaked period is shown to be around 5.9 hours with
associated flanking peaks from 24 hours and 15 and 60 day sampling
aliases (Figure~\ref{fig:krpdm}).  Closer examination of the PDM fit
near 5.9 hours shows the 15 and 60 day aliasing much better and gives
two best fit periods, one at 5.840 and the other at 5.929 hours
(Figure~\ref{fig:krpdmsmall}).  We phased the 2001 data to both single
peaks and found neither to be significantly better than the other.
The true single-peaked period for 1999 KR$_{16}$ is at one of these
two values.  The data phased to the $5.840$ hour single-peaked period
are shown in Figure~\ref{fig:lightkr1}. Neither of the possible
double-peaked periods of $11.680$ and $11.858$ hours show differences
between the peaks.  The peak-to-peak amplitude of 1999 KR$_{16}$ is
$0.18 \pm 0.04$ in the 2001 data consistent with that found in the
2000 data.  Colors of 1999 KR$_{16}$, $B-V=0.99\pm 0.05$, $V-R=0.75\pm
0.04$, and $R-I=0.70\pm 0.04$, are on the red end of the KBO
distribution (Table~5).  The colors show no signs of variation through
the rotation of the object to the accuracy of our measurements
(Table~8 and Figure~\ref{fig:lightkr1}).

\subsection{Null Lightcurves}
Nine of the TNOs (2001 FZ$_{173}$, 2001 CZ$_{31}$, (38628) 2000
EB$_{173}$, (26375) 1999 DE$_{9}$, 1998 HK$_{151}$, (33340) 1998
VG$_{44}$, (19521) Chaos 1998 WH$_{24}$, 1997 CS$_{29}$, and (26181)
1996 GQ$_{21}$) show no measurable photometric
variations. Practically, this means that their lightcurves have range
$\leq 0.15$ magnitudes and/or period $\geq 24$ hours
(Figures~\ref{fig:lightall} and Table~4).  A few objects show hints of
variability that might, with better data, emerge as rotationally
modulated lightcurves.  Inspection of the 2001 CZ$_{31}$ data hints at
a single-peaked lightcurve of period $\sim 3$ hours and amplitude
$\sim 0.15$ magnitudes, but since the photometry has large error bars
we can not be sure of this result.  The TNO 1999 DE$_{9}$ may have a
long period lightcurve of about 0.1 mag. range since the brightness on
2001 April 24 slowly increases towards the end of the night and the
February data appear to have base magnitudes different by about 0.1
mag.  The data from 2000 on 1999 DE$_{9}$ show the object to have a
flat lightcurve.  (33340) 1998 VG44 may also have a long period
lightcurve since its base magnitudes on 1999 November 11 and 12 are
different by about 0.05 mag. The bright TNO (19521) 1998 WH$_{24}$ may
have a possible lightcurve of about 4 hours single-peaked period and
peak-to-peak range of 0.07 mag.  Confirmation of these subtle
lightcurves will require more accurate data, probably from larger
telescopes than the one employed here.

\section{Interpretation}

The KBOs should be in principal axis rotation since the expected
damping time of any other wobbles is much less than the age of the
Solar System (Burns \& Safronov 1973; Harris 1994).  Orbital periods
of KBOs are long ($>200$ years) and thus the pole orientation to our
line of sight should not change significantly between epochs.  The
apparent magnitude of a KBO depends on its physical characteristics
and geometrical circumstances and can be represented as
\begin{equation}
m_{R}=m_{\odot}-2.5\mbox{log}\left[p_{R}r^{2}\phi (\alpha )/(2.25\times 10^{16}R^{2}\Delta^{2})\right]  \label{eq:appmag}
\end{equation}
in which $r$ [km] is the radius of the KBO, $R$ [AU] is the
heliocentric distance, $\Delta$ [AU] is the geocentric distance,
$m_{\odot}$ is the apparent red magnitude of the sun ($-27.1$),
$m_{R}$ is the apparent red magnitude, $p_{R}$ is the red geometric
albedo, and $\phi (\alpha)$ is the phase function in which the phase
angle $\alpha=0$ deg at opposition and $\phi (0)=1$.  The apparent
brightness of an inert body viewed in reflected light may vary because
of 1) changes in the observing geometry, including the effects of
phase darkening as in Eq. (1) and 2) rotational modulation of the
scattered light.  These different effects are discussed below.

\subsection{Non-uniform Surface Markings}

Surface albedo markings or topographical shadowing can potentially
influence the lightcurves.  Judging by other planetary bodies, the
resulting light variations are typically smaller than those caused by
elongated shape, with fluctuations due to albedo being mostly less
than about 10 to 20 percent (Degewij, Tedesco, Zellner 1979).  A color
variation at the maximum and minimum of a lightcurve may be seen if
albedo is the primary cause for the lightcurve since materials with
markedly different albedos often also have markedly different colors.
For example, many pure ices and frosts have a very high albedo and are
neutral to bluish in color. A lightcurve caused by an ice or frost
patch should show a bluish color when at maximum brightness.  Some of
the most extreme albedo contrasts are found on Pluto and the Saturnian
satellite Iapetus (Table~9).  The latter is in synchronous rotation
around Saturn with its leading hemisphere covered in a very low albedo
material thought to be deposited from elsewhere in the Saturn
system. Iapetus shows clear rotational color variations ($\Delta(B-V)
\sim$ 0.1 mag.)  that are correlated with the rotational albedo
variations.  On the other hand, Pluto has large albedo differences
across its surface but the hemispherically averaged color variations
are only of order 0.01 mag.  We feel that neither Iapetus nor Pluto
constitutes a particularly good model for the KBOs.  The large albedo
contrast on Iapetus is a special consequence of its synchronous
rotation and the impact of material trapped in orbit about Saturn.
This process is without analog in the Kuiper Belt. Pluto is also not
representative of the other KBOs.  It is so large that it can sustain
an atmosphere which may contribute to amplifying its lightcurve
amplitude by allowing surface frosts to condense on brighter (cooler)
spots.  Thus brighter spots grow brighter while darker (hotter) spots
grow darker through the sublimation of ices.  This positive feedback
mechanism requires an atmosphere and is unlikely to be relevant on the
smaller KBOs studied here.

\subsection{Aspherical Shape}
The critical rotation period ($T_{crit}$) at which centripetal acceleration
equals gravitational acceleration towards the center of a rotating
spherical object is

\begin{equation}
T_{crit} = \left(\frac{3\pi }{G \rho}\right)^{1/2}   \label{eq:equil}
\end{equation}

where $G$ is the gravitational constant and $\rho$ is the density of
the object.  With $\rho$ = $10^3$ kg m$^{-3}$ the critical period is
about 3.3 hours.  Even at longer periods, real bodies will suffer
centripetal deformation into aspherical shapes.  For a given density
and specific angular momentum (H), the nature of the deformation
depends on the strength of the object. In the limiting case of a
strengthless (fluid) body, the equilibrium shapes have been well
studied (Chandrasekhar 1987).  For $H \leq$ 0.304 (in units of $(GM^{3}
a^{'})^{1/2}$, where $M$[kg] is the mass of the object and $a^{'}$[m]
is the radius of an equal volume sphere) the equilibrium shapes are the
oblate "MacLaurin" spheroids.  Oblate spheroids in rotation about their
minor axis exhibit no rotational modulation of the cross-section and
therefore are not candidate shapes for explaining the lightcurves of
the KBOs.  However, for 0.304 $\leq H \leq$ 0.390 the equilibrium
figures are triaxial "Jacobi" ellipsoids which generate lightcurves of
substantial amplitude when viewed equatorially.  Strengthless objects
with $H >$ 0.390 are rotationally unstable to fission.

The KBOs, being composed of solid matter, clearly cannot be
strengthless.  However, it is likely that the interior structures of
these bodies have been repeatedly fractured by impact, and that their
mechanical response to applied rotational stress is approximately
fluid-like. Such ``rubble pile'' structure has long been suspected in
the main asteroid belt (Farinella et al. 1981) and has been
specifically proposed to explain the short period and large amplitude
of (20000) Varuna (Jewitt and Sheppard 2002).  The rotational
deformation of a rubble pile is uniquely related to its bulk density
and specific angular momentum.  Therefore, given that the shape and
specific angular momentum can be estimated from the amplitude and
period of the lightcurve, it is possible to use photometric data to
estimate the density.

Elongated Objects exhibit rotational photometric variations caused by
changes in the projected cross-section.  The rotation period of an
elongated object should be twice the single-peaked lightcurve
variation due to its projection of both long axes (2 maxima) and short
axes (2 minima) during one full rotation.  From the ratio of maximum
to minimum brightness we can determine the projection of the body
shape into the plane of the sky.  The rotational brightness range of a
triaxial object with semiaxes $a \geq b \geq c$ in rotation about the
$c$ axis is given by (Binzel et al. 1989)

\begin{equation}
\Delta m=2.5\mbox{log}\left(\frac{a}{b}\right) - 1.25\mbox{log}\left(\frac{a^{2}cos^{2}\theta +c^{2}sin^{2}\theta}{b^{2}cos^{2}\theta +c^{2}sin^{2}\theta}\right)   
\label{eq:elong}
\end{equation}

where $\Delta m$ is expressed in magnitudes, and $\theta$ is the angle
at which the rotation ($c$) axis is inclined to the line of sight (an
object with $\theta = 90$ deg. is viewed equatorially).

It is to be expected that, through collisions, fragments would have
random pole vector orientations.  For example, the collisionally
highly evolved asteroid belt shows a complete randomization of pole
vector orientations, $\theta$. Only the largest asteroids may show a
preference for rotation vectors aligned perpendicular to the ecliptic
($\theta = 90\arcdeg$), though this is debatable (Binzel et al. 1989;
Drummond et al. 1991; De Angelis 1995).  In the absence of any pole
orientation data for the KBOs, we will assume they have a random
distribution of spin vectors.  Given a random distribution, the
probability of viewing an object within the angle range $\theta$ to
$\theta + d\theta$ is proportional to sin$(\theta) d\theta$.  In such
a distribution, the average viewing angle is $ \theta = 60$ degrees.
Therefore, on average, the sky-plane ratio of the axes of an elongated
body is smaller than the actual ratio by a factor sin$(60) \approx
0.87$.

In addition to rotational deformation, it is possible that some
asteroids and KBOs consist of contact binaries (Jewitt \& Sheppard
2002).  For a contact binary consisting of equal spheres, the axis
ratio of 2:1 corresponds to a peak-to-peak lightcurve range $\Delta m$
= 0.75 mag., as seen from the rotational equator.  For such an object
at the average viewing angle $\theta = 60$ degrees we expect $\Delta
m$ = 0.45 mag.

Collisionally produced fragments on average have axis ratios
$2:2^{1/2}:1$ (Fujiwara, Kamimoto, \& Tsukamoto 1978; Capaccioni et
al.  1984). When viewed equatorially, such fragments will have $\Delta
m$ = 0.38 mag.  At the mean viewing angle $\theta = 60$ degrees we
obtain $\Delta m$ = 0.20 mag.

\subsection{Lightcurve Model Results}

The KBOs in our sample are very large ($D>250$ km assuming a low
albedo) and should, in the absence of rotational deformation, be
spherical in shape from gravitational self compression.  The large
amplitudes and fast rotations of (20000) Varuna, 2000 GN$_{171}$, and
(33128) 1998 BU$_{48}$ suggest that the lightcurves are caused by
elongation and not surface albedo features.  In support of this is the
finding that (33128) 1998 BU$_{48}$ and (20000) Varuna have no color
variations throughout their lightcurves and 2000 GN$_{171}$ has only a
slight if any variation in color.  Independently 2000 GN$_{171}$ shows
two distinct lightcurve maxima and minima which is a strong reason to
believe the object is elongated.  The other lightcurve we found was
for 1999 KR$_{16}$.  Since its amplitude is much smaller and period
longer, the lightcurve of 1999 KR$_{16}$ may be more dominated by
nonuniform albedo features on its surface, though we found no
measurable color variation over the rotation.

Table~10 lists the parameters of albedo, Jacobi ellipsoid and binary
models that fit the axis ratios estimated from the lightcurve data
(Table~4).  For each object and model, we list the minimum bulk
density, $\rho$, required to maintain a stable configuration, as
described in Jewitt and Sheppard (2002).  We briefly describe the
procedure below for 2000 GN$_{171}$.  Results for the rest of the
significant light variation objects in our sample ((20000) Varuna,
(33128) 1998 BU$_{48}$, and 1999 KR$_{16}$) can be seen in Table~10
using the data from Table~4.

We use Equation~\ref{eq:elong} to estimate the axis ratio $a/b$.  If
we assume that the rotation axis is perpendicular to our line of sight
($\theta =90$) we obtain

\begin{equation}
\frac{a}{b} = 10^{0.4\Delta m_R}
\label{eq:axisratio}
\end{equation}

Using $\Delta m_R$ = 0.61 magnitudes we obtain from
Equation~\ref{eq:axisratio} $a/b$ = $1.75:1$ for 2000 GN$_{171}$.
This is a lower limit to the intrinsic axis ratio because of the
effects of projection into the plane of the sky.  If 2000 GN$_{171}$
is a Jacobi triaxial ellipsoid with $P=8.329$ hours then its $a:b:c$
axis ratio would be $1.75:1:0.735$ and the lower limit on the density
would be $\rho = 635$ kg m$^{-3}$ (Chandrasekhar 1987; see Jewitt \&
Sheppard 2002 for a KBO context discussion of Jacobi ellipsoids).
Finally if 2000 GN$_{171}$ were a contact binary the ratio of the two
radii, $a_{1}:a_{2}$, would be $1.15:1$ with a lower limit to the
density of $\rho =585$ kg m$^{-3}$ (see Jewitt \& Sheppard 2002 for a
discussion of contact binaries in the KBO context).  Finally, though
it is unlikely, if 2000 GN$_{171}$ is spherical and the lightcurve is
due to a $1.75:1$ contrast in albedo then the lower limit to the
density of the KBO would be $\rho = 157$ kg m$^{-3}$ from
Equation~\ref{eq:equil} and using $P=8.329$ hours.

\section{Discussion}

In Table~9 we show objects in the Solar System which have one axis of
at least 200 km and which show large amplitude lightcurves.
Interestingly there is a group of asteroids that are large ($D=200$ to
$300$ km) and which have substantial lightcurve amplitudes.  They also
possess fast rotations.  These objects are probably rotationally
deformed ``rubble piles'' which may be similar to a Jocabi ellipsoid
type object (Farinella et al.  1981).  Such rubble pile structures may
form in the main asteroid belt because all objects have been effected
by the high-velocity ($\sim 5$ km/s) collisions that occur there
(Farinella, Paolicchi, Zappala 1982). The effect of collisions is
highly dependent on the object size.  Objects with $D > 300$ km are
large enough not to be completely turned into rubble piles or have
their momentum greatly altered.  Objects with diameters $200$ to $300$
km are large enough to be gravitationally bound but impacts over the
age of the Solar System will transform them into rubble piles and may
significantly change their angular momentum.  Most asteroids with $D<
200$ km are thought to be fragments from catastrophic collisions and
are not massive enough to be gravitationally spherical.

How does the collisional outcome scale with velocity and density
differences in the asteroid belt versus the Kuiper Belt?  We assume
the target body has catastrophic break up when the projectile kinetic
energy equals the gravitational binding energy of the target

\begin{equation}
\frac{1}{2} M_{p} \Delta v^{2} = \frac{3G M_{t}^{2}}{5r_{t}}
\label{eq:binding}
\end{equation}

where $\Delta v$ is the mean collisional speed, $M$ is mass, $r$ is
radius, and subscripts $p$ and $t$ refer to projectile and target,
respectively.  For collisions with a target of given radius, the ratio
of the sizes of the projectiles needed to cause disruption in the
main-belt and in the Kuiper Belt is

\begin{equation}
\frac{r_{p,KB}}{r_{p,MB}} =\left[\left(\frac{\rho_{t,MB}}{\rho_{t,KB}}\right) \left(\frac{\Delta v_{KB}}{\Delta v_{MB}}\right)^{2}\right]^{-1/3} 
\label{eq:binding2}
\end{equation}

where we have assumed all Kuiper Belt objects have density
$\rho_{KB}$, all main belt asteroids have density $\rho_{MB}$.  Here
$r_{p,MB}$ and $r_{p,KB}$ are the radii of the projectile in the main
belt and Kuiper Belt which are needed to fracture the target in their
respective belts, $\rho_{t,MB}$ and $\rho_{t,KB}$ are the densities of
the target body in the main belt and Kuiper Belt respectively, and
$\Delta v_{MB}$ and $\Delta v_{KB}$ are the respective collision
velocities.  If we put in nominal values of $\rho_{t,MB} = 3000$ kg
m$^{3}$, $v_{MB} = 5$ km s$^{-1}$ and $\rho_{t,KB} = 1000$ kg m$^{3}$,
$v_{KB} = 1.5$ km s$^{-1}$ for the main belt asteroids and Kuiper Belt
respectively we find
\begin{equation}
r_{p,KB} \approx 1.5  r_{p,MB}.
\label{eq:binding3}
\end{equation}
Thus for targets of equal size, a projectile has to be about $50\%$
larger in the Kuiper Belt than in the main belt to be able to cause
catastrophic break up of the target body.  This difference is not
large and since the current collisional timescales for the asteroids
and Kuiper Belt objects are similar (Davis \& Farinella 1997; Durda \&
Stern 2000), other factors such as material strength and the number
density of objects during early formation of each belt will be
important in determining collisional differences.

The current Kuiper Belt has been found to be erosive for KBOs with
$D<100$ km while many of the larger objects are probably rubble piles
(Davis \& Farinella 1997).  Laboratory and computer simulations show
that self-gravitating targets are more easily fractured than dispersed
(Asphaug et al. 1998).  Once formed, rubble pile structures can
insulate the rest of the body from the energy of impact, further
inhibiting disruption.  Collision experiments by Ryan, Davis, and
Giblin (1999) also show that porous ices dissipate energy efficiently.
The outcome of impact into a rubble pile depends heavily on the angle
of impact.  We note that glancing low velocity collisions
substantially alter the spin of the target body and can create
elongated objects and contact binaries (Leinhardt, Richardson, \&
Quinn 2000).  These simulations all hint that rubble pile structures
are able to remain gravitationally bound after an impact, but that
their angular momentum may be altered in the process which could
produce elongated shapes.

To date eight binary Kuiper Belt objects have been reported.  It seems
that there may be a large fraction of binary KBOs.  It also appears
that about $32 \%$ of KBOs are highly elongated.  Both the binaries
and the highly elongated shapes indicate large specific angular
momentum, most likely delivered by glancing collisions.  The current
rate of collisions is too small however for any substantial
modifications of the spins or shapes of KBOs (Jewitt and Sheppard
2002).  Instead, we prefer the hypothesis that the binaries and
elongated shapes are products of an early, denser phase in the Kuiper
Belt, perhaps associated with its formation.

\subsection{Other Lightcurve Observations}

We now consider lightcurve observations of KBOs published by others in
order to make a larger sample.  Unfortunately, few KBOs to date have
been shown through independent observations to have repeatable
lightcurves.  Hainaut et al. (2000) reported that (19308) 1996
TO$_{66}$ has a lightcurve which varies in amplitude over the course
of one year and interpreted this as a result of possible on-going
cometary activity.  Object 1996 TO$_{66}$ may show a color difference
throughout its rotation (Sekiguchi et al. 2002).  In contrast, 1996
TO$_{66}$ was reported to have a flat lightcurve by Romanishin \&
Tegler (1999) during the same year in which Hainaut et al. (2000)
detected variation.  Our own observations show that 1996 TO$_{66}$
does have a significant lightcurve, basically confirming the variation
originally observed by Hainaut et al. (2000) and contradicting the
null detection by Romanishin \& Tegler (Sheppard 2002).  Conversely,
an object reported to have a lightcurve by Romanishin \& Tegler
(1999), (15820) 1994 TB, was found by us to display no significant
variation (Sheppard 2002).  Because of these conflicts of
unrepeatability, and since many of the Romanishin \& Tegler targets
were very sparsely sampled with raw data that remains unpublished, we
use their work with caution in the following analysis.

Our combined sample of 22 KBOs comprises only well observed objects
with numerous observations that could constrain any significant
photometric variation from this (Table~4) and other (Table~11)
works. Among the objects newly observed in this survey (Table~4), the
fraction with significant lightcurve variation is $f(\Delta m_R \geq
0.15) = \frac{4}{13}$ ($31\%$) and $f(\Delta m_R \geq 0.40) =
\frac{3}{13}$ ($23\%$). Including the objects reliably observed by
others (Table~11) yields $f(\Delta m_R \geq 0.15) = \frac{7}{22}$
($32\%$) and $f(\Delta m_R \geq 0.40) = \frac{5}{22}$ ($23\%$).
Although we have evidence that some of their lightcurves are
unrepeatable, we note that Romanishin \& Tegler (1999) found a
comparable $f(\Delta m_R \geq 0.10) = \frac{3}{11}$ ($27\%$).  We
consider that these results all point to a similar fraction $f(\Delta
m_R \geq 0.15) \sim$ $32\%$ and $f(\Delta m_R \geq 0.40) \sim$ $23\%$.

The samples of objects with significant lightcurves and flat
lightcurves were tested for correlations with orbital parameters and
colors.  No significant correlations were found.  From the sample of
22 objects, 2 of the 9 ($22\%$) resonant objects, 4 of the 8 ($50\%$)
classical objects, and 1 of the 5 ($20\%$) scattered objects had
measurable lightcurves ($\Delta m_{R}\ge 0.15$).  Many of the objects
shown in Table~11 are detailed elsewhere by us (Sheppard 2002) because
they were objects particularly targeted by us to confirm their
reported lightcurves and determine amplitudes and periods if a
lightcurve was seen.  The 13 objects reported in this paper (Table~4)
were selected because of their size and brightness and not because of
previous reports of their variability.

In comparison to the percentages of KBOs with large amplitude
lightcurves ($>0.40$ or about 1.5 difference in brightness), the four
main belt asteroids with $D>400$ km have $f(\Delta m_R \geq 0.40) =
\frac{0}{4}$ ($0\%$), the largest being only about 0.15 magnitudes
(Lagerkvist, Harris, \& Zappala 1989; Tedesco 1989).  For main-belt
asteroids with $D > 200$ km $f(\Delta m_R \geq 0.40) = \frac{5}{27}$
($19\%$) when their poles orientations are $\theta = 90$ degrees to
our line of site.  With the average pole orientation of $\theta = 60$
degrees only ($11\%$) ($f(\Delta m_R \geq 0.40) = \frac{3}{27}$) have
large amplitude lightcurves.  These large amplitude lightcurve objects
are thought to be the Jacobi ellipsoid type objects.  Asteroids with
$D<200$ km have $f(\Delta m_R \geq 0.40) = \frac{111}{482}$ ($23\%$)
while the Centaurs (Chiron, Asbolus, Pholus, Chariklo, Hylonome,
(31824) 1999 UG5, and (32532) 2001 PT13) have $f(\Delta m_R \geq 0.40)
= \frac{0}{7}$ ($0\%$).  These objects are small and thus thought to
be collisional fragments.

Figure~\ref{fig:ampper} shows how the largest ($D>200$ km) main belt
asteroids compare with the Kuiper Belt objects.  Many of the Kuiper
Belt objects fall in the upper and upper left parts of this figure,
where the Jacobi ellipsoids are encountered in the asteroid belt.
There is a bias in the KBO sample since light variations of less than
about 0.1 magnitudes are very hard to detect, as are long single-peak
periods $>24$ hours.

The Student's t-test was used to measure the significance of the
differences between the means of the asteroid and KBO periods and
amplitudes.  In order to reduce the effects of observational bias we
used only periods less than 10 hours and amplitudes greater than 0.2
magnitudes from Figure~\ref{fig:ampper}.  We found that the period
distributions of the asteroids are significantly shorter than for the
KBOs.  The mean periods are $5.56\pm 0.89$ and $7.80\pm 1.20$ hours
for the asteroids and KBOs respectively, giving a t-statistic of
$-3.84$ (12 degrees of freedom) which is significant at the 99.7\%
confidence level.  This difference is formally significant at the
$3\sigma$ level by the Student's t-test, but it would be highly
desirable to obtain more data from another large unbiassed survey in
order to be sure of the effect.  The KBOs have a larger mean
amplitude, but the significance between the difference of means,
$0.36\pm 0.11$ vs. $0.50\pm 0.16$ magnitudes for the asteroids and
KBOs respectively, is only 95\% ($2\sigma$) with a t-statistic of
$-1.83$.  This may be because the KBOs are less dense and more
elongated, on average, than asteroids.  Below we discuss in more
detail the shape distribution of the Kuiper Belt.

\subsection{Shape Distribution Models}

What constraints can be placed on the intrinsic distribution of KBO
shapes from the apparent (sky-plane projected) distribution?  We used
a Monte-Carlo model to project several assumed intrinsic distributions
into the plane of the sky and then compared them with the
observations.  This was done by using a pole orientation distribution
proportional to sin$\theta$.  The apparent axis ratio for each object
was then calculated from this pole orientation distribution and the
intrinsic axis ratio selected from one of several assumed
distributions.

Firstly, as an extreme case, we ask whether the data are consistent
with selection from intrinsic distributions in which all the objects
have a single axis ratio $x=b/a$, with $x = 0.80,0.66,0.57$ or $0.50$
(Figure~\ref{fig:simulation1}).  The Figure shows that the form of the
resulting amplitude distribution differs dramatically from what is
observed.  We conclude that the distribution KBO lightcurve amplitudes
cannot be modeled as the result of projection on any single axis
ratio.  A range of shapes must be present.  While not surprising, this
result does serve to demonstrate that the KBO lightcurve sample is of
sufficient size to be diagnostic.

Secondly, we explored the effect of the width of the distribution using 
\begin{equation}
\Psi(x)dx= exp\left[\frac{-(x - x_{0})^2}{2\sigma^2}\right] dx
\label{eq:simmodel3}
\end{equation}
where $\Psi(x)dx$ is the number of KBOs with axis ratios in the range
$x$ to $x + dx$, $\sigma$ is the standard deviation or width parameter
and $x_{0}$ is the mean axis ratio.  Examples for $x_{0}$ = 0.66 and
$\sigma$ = 0, 0.35, 0.75, and 1.0 are plotted in Figure
~\ref{fig:simulation2a}.  We assumed that all objects had axis ratios
$0.5 \leq x \leq 1.0$.  The Figure shows that the data require an
intrinsically broad distribution of body shapes, specifically with a
dispersion comparable to the mean axis ratio.

Thirdly, we assumed that the axis ratios of the KBOs followed a
differential power-law distribution of the form
\begin{equation}
\Psi(x)dx = x^{-q} dx
\label{eq:simmodel4}
\end{equation}
where $q$ is a constant, and $\Psi(x)dx$ is again the number of KBOs
with axis ratios in the range $x$ to $x + dx$.  We assumed $0.5 \leq x
\leq 1.0$.  A positive $q$ favors objects with small axis ratios while
negative $q$ favors objects that are near spherical.  The results can
be seen in Figure~\ref{fig:simulation2b}.  The $q=-5$ distribution is
very similar to an exponential distribution with its peak at an axis
ratio of $x=1$.  Again we see that the models fit the data better with
a broader distribution of axis ratios.

Fourthly, we ask whether the data are consistent with selection from
an intrinsic distribution of shapes caused by collisional
fragmentation.  The fragment shape distribution is taken from Catullo
et al. (1984).  Figure~\ref{fig:simulation3} shows that the KBO
$\Delta m$ distribution is inconsistent with the collisional fragment
distribution in the sense that more highly elongated KBOs are found
than would be expected from the impact fragments.  This finding is
consistent with collisional models (Farinella and Davis 1996, Kenyon
and Luu 1999) in the sense that only KBOs smaller than a critical
diameter $\sim$ 100 km are likely to be impact fragments, while the
observed KBOs are all larger than this.

Finally, we ask whether the data are consistent with selection from an
intrinsic distribution of shapes like that measured in the large (D $>
200$ km) main-belt asteroid population.  We take this distribution
from the published lightcurve data base of Lagerkvist, Harris, \&
Zappala (1989) which has been updated by A. Harris on the world wide
web at http://cfa-www.harvard.edu/iau/lists/LightcurveDat.html.  The
results are shown in Figure~\ref{fig:simulation3}, where we see that
the KBOs contain a larger fraction of highly elongated objects than
are found amongst the main-belt asteroids.  A plausible explanation
for such a large fraction of the highly elongated Kuiper Belt objects
is that the objects are very large yet structurally weak and of low
density.  This would allow many of the Kuiper Belt objects to be
gravitationally bound rubble piles easily distorted by centripetal
forces due to their rotation.

\subsection{KBO Density Comparisons in the Solar System}

The Kuiper Belt objects are thought to consist of water ice with some
rocky material mixed in, similar to the comets.  How do the densities
of the outer satellites compare to what we have found for our sample
of Kuiper Belt objects?  In Figure~\ref{fig:iceden} we plot all the
outer icy bodies in the Solar System that have well known densities
and are less than 3000 km in diameter.  There is a clear trend, with
larger objects being denser.  The KBOs seem to follow this trend.  We
also note there appears to be an object size vs. lightcurve amplitude
and size vs. period trend for the KBOs in our data.  Objects that have
densities less than that of water ice (1000 kg m$^{-3}$) must have
significant internal porosity or be composed of ices less dense than
water (see Jewitt and Sheppard 2002).

To date only about 10 main belt asteroids have reliably measured bulk
densities.  Most of these are from perturbation calculations between
asteroids though two have been measured by passing spacecraft and a
few others found from the orbital motions of known companions.  Most
asteroid densities are consistent with that of rock, $2000 \leq \rho
\leq 3000$ kg m$^{-3}$.  Some of the asteroids densities have been
found to be lower than expected and attributed to internal porosity
possibly from rubble pile structure (Yeomans et al. 1997).

In Table~9 we present new densities for five main belt asteroids
calculated under the assumption that they are equilibrium rotational
(Jacobi ellipsoid) figures.  We used their lightcurves as seen at
maximum amplitude, to eliminate the effects of projection.  The
densities are higher than those of the Kuiper Belt objects obtained
using the same method (Figure~\ref{fig:iceast}) but lower than
expected for solid rock objects.  This provides another hint that
these objects may be internally porous.  The densities of 15 Eunomia
($790\pm 210$ kg m$^{-3}$) and 16 Psyche (1800$\pm 600$ kg m$^{-3}$)
were reported separately from measurements of gravitational
perturbations (Hilton 1997; Viateau 2000).  The higher density for 16
Psyche is particularly interesting because this object is an M-type
asteroid and thus expected to have a high density.  The main belt
asteroid 45 Eugenia was found to have a companion which was used by
Merline et al. (1999) to find a density of $1200_{-200}^{+600}$ kg
m$^{-3}$.  Asteroid densities found by others are probably
underestimated since they assumed that the objects were spheres.  A
sphere has the highest volume to projected area ratio and thus any
deviation from a sphere will cause the object to appear to have a
lower density.  We calculated the density for these objects using the
assumption they are Jacobi ellipsoids and thus the parameters used are
the well known period and amplitude from the lightcurves.
Interestingly the five best examples of main belt rotationally
deformed asteroids (Table~9) are found in all the main classes, 2
C-type, 1 each of S, P, and M-types.

\section{Phase Functions of KBOs}

At large phase angles, the phase function in Equation~\ref{eq:appmag}
may be approximated as
\begin{equation} 
\phi(\alpha) = 10^{-\beta \alpha}
\label{eq:phangle}
\end{equation} 
where $\alpha$ is the phase angle in degrees, and $\beta$ is the
"linear" phase coefficient.  Empirically, the magnitude of $\beta$ is
inversely correlated with the surface albedo (Gehrels 1970; Bowell et
al. 1989; Belskaya and Shevchenko 2000), suggesting that we might be
able to indirectly assess the albedos of KBOs from their phase
functions.  Unfortunately, this is not possible.  The maximum phase
angle attained by an object at distance $R$ [AU] is roughly
$\alpha_{max}$ [degrees] = $\frac{180}{\pi R}$.  At $R = 30$ AU, for
instance, $\alpha_{max} =$ 1.9 degrees.  This is exactly the phase
angle range in which the opposition surge is potentially important
(Scaltriti and Zappala 1980; Belskaya and Shevchenko 2000).  The
opposition surge is a complex, multiple scattering phenomenon which
occurs in the grains of a porous regolith.  The magnitude of the
opposition surge, which causes an increase in scattered intensity over
and above that predicted by Equation~\ref{eq:phangle} at small
$\alpha$, is determined by coherent-backscattering and is a complex
function of regolith physical and optical properties.  It is not
simply related to the albedo and Equation~\ref{eq:phangle} must be
modified to take account of this surge.  Nevertheless, the phase
functions provide a new basis for comparison of the KBOs, and should
be measured if we are to accurately assess the sizes of KBOs from
their optical data.

Seven of the KBOs were observed over a range of phase angles
sufficient for us to measure the phase darkening.  We plot the
quantity $m_{R}(1,1,\alpha) = m_{R} - 5\mbox{log}(R\Delta)$ against
$\alpha$ for these 7 KBOs in Figures~\ref{fig:phaseall} and
Figure~\ref{fig:phaseallG}.  When observations from consecutive nights
were available we averaged the phase angle and apparent magnitude over
those nights to create a single point with small uncertainty.  If an
object showed a lightcurve, its time-averaged mean apparent magnitude
was used.  The linear least squares fits to the KBO data are listed in
Table~12 and shown in Figure~\ref{fig:phaseall}.  Within the
uncertainties, we find that photometry of the 7 KBOs is compatible
with $\beta (\alpha <2\arcdeg)$ = $0.15\pm 0.01$ mag deg$^{-1}$.  In
contrast the phase function for Pluto was found to be linear
throughout the 0 to 2 degrees phase angle range with $\beta (\alpha
<2\arcdeg) = 0.0372\pm 0.0016$ mag deg$^{-1}$, indicating a very
shallow if any opposition surge and consistent with a high albedo
surface (Tholen and Tedesco 1994).

Since the small phase angle observations are affected by the
"opposition surge", caused by multiple scattering within the porous
regolith, we also fit the data using the Bowell et al. (1989) $H-G$
scattering parametrization.  This technique gives a curved relation at
small phase angles that becomes asymptotically like the linear $\beta$
relation at large phase angles and thus attempts to account for the
opposition surge.  In the Bowell et al. formalism $H$ is the absolute
magnitude of the object, analogous to $m_R(1,1,0)$.  The parameter $G$
provides a measure of the slope of the phase function at large angles,
analogous to $\beta$. It is scaled so that $G=0$ corresponds to the
darkest surfaces found on the asteroids while $G=1$ corresponds to the
brightest (Bowell et al. 1989).  The results of the $H-G$ fits are
presented in Table~12 and Figures~\ref{fig:phaseallG} and
~\ref{fig:phasekr}. The KBOs show steep slopes with a possible
moderate opposition surge.  The best-fit values of the $G$ parameter
are very low with an average of $-0.21$.  This small $G$ value more
closely resembles that of dark, C-type asteroids ($G\sim 0.15$) than
the brighter, S-types ($G\sim 0.25$) in the main-belt.  This is
consistent with, though does not prove, the assumption that the
majority of KBOs are of very low albedo.  The similarity of the slopes
of the phase functions of all KBOs in our sample suggests comparative
uniformity of the surface compositions, physical states, and albedos.
As a comparison, Pluto was found to have a best fit $G = 0.88\pm 0.02$
using data from Tholen \& Tedesco (1994).  The dramatic difference
between the backscattering phase functions of Pluto and the smaller
KBOs studied here is shown in Figure~\ref{fig:phasekr}.  This
difference is again consistent with the smaller KBOs having low albedo
(0.04?) surfaces qualitatively different from the high albedo (0.6),
ice-covered surface of Pluto.

\section{Summary}

We have conducted a systematic program to assess the rotations and sky-plane
shapes of Kuiper Belt Objects from their optical lightcurves.  

1. Four of 13 ($31 \%$) bright Kuiper Belt objects in our sample (
(33128) 1998 BU$_{48}$, 2000 GN$_{171}$, (20000) Varuna, and 1999
KR$_{16}$) show lightcurves with range $\Delta m \geq$ 0.15 mag. In an
enlarged sample combining objects from the present work with objects
from the literature, 7 of 22 ($32 \%$) objects have $\Delta m \geq$
0.15 mag.

2. The fraction of KBOs with $\Delta m \geq$ 0.4 mag ($23 \%$) exceeds
the corresponding fraction in the main-belt asteroids ($11 \%$) by a
factor of two.  The KBO $\Delta m$ distribution is inconsistent with
the distribution of impact fragment shapes reported by Catullo et
al. (1984).

3. The large Kuiper Belt Objects (33128) 1998 BU$_{48}$, 2000
GN$_{171}$ and (20000) Varuna show large periodic variability with
photometric ranges $0.68\pm 0.04$, $0.61\pm 0.03$ and $0.45\pm 0.03$
magnitudes, respectively, and short double-peaked periods of $9.8\pm
0.1$, $8.329\pm 0.005$ and $6.3565\pm 0.0002$ hours, respectively.
Their BVRI colors are invariant with respect to rotational phase at
the few percent level of accuracy.

4. If these objects are equilibrium rubble piles distorted by
centripetal forces due to their own rotation, the implied densities
must be comparable to or less than that of water.  Such low densities
may be naturally explained if the KBOs are internally porous.

5. In the phase angle range $0 \leq \alpha \leq 2$ deg the average
slope of the phase function of 7 KBOs is $\beta (\alpha < 2\arcdeg)$ =
$0.15\pm 0.01$ mag deg$^{-1}$ (equivalently, $G=-0.2$).  The
corresponding slope for ice-covered Pluto is $\beta (\alpha < 2\arcdeg
) \approx 0.04$ mag/deg (equivalently, $G=0.88$).  The large
difference is caused by pronounced opposition brightening of the KBOs,
strongly suggesting that they possess porous, low albedo surfaces
unlike that of ice-covered Pluto.

\section*{Acknowledgments}

We thank John Dvorak, Paul deGrood, Ian Renaud-Kim, and Susan Parker
for their operation of the UH telescope, Alan Harris for a quick and
thoughtful review.  This work was supported by a grant to D.J. from
NASA.

\newpage

\begin{figure}
\caption{The phase dispersion minimization (PDM) plot for 2000 GN$_{171}$.
A smaller theta corresponds to a better fit.  Best fits from this plot
are the 4.12 hour single-peaked fit and the 8.32 hour double-peaked
fit.  Both are flanked by 24 hour alias periods.} 
\label{fig:gnpdm} 
\end{figure}

\begin{figure}
\caption{Phased R-band data from the UT April $20-25$ and May $11-13$,
2001 observations of 2000 GN$_{171}$.  The period has been phased to
4.17 hours which is the best fit single-peaked period.  The May data
have been corrected for geometry and phase angle differences relative
to the April data (see Table~1).  The points are much more scattered
here than for the better fit double-peaked period
(Figure~\ref{fig:lightgnr}).}
\label{fig:lightgnsingler}
\end{figure}

\begin{figure}
\caption{Phased R-band data from the UT April $20-25$ and May $11-13$,
2001 observations for 2000 GN$_{171}$.  The period has been phased to
8.329 hours which is the best fit double-peaked period.  The May data
have been corrected for geometry and phase angle differences relative
to the April data (see Table~1). }
\label{fig:lightgnr} 
\end{figure}

\begin{figure}
\caption{Closer view of the phase dispersion minimization (PDM) plot
for 2000 GN$_{171}$ around the doubled-peaked period near 8.33 hours.
The best fit at 8.329 hours is flanked by aliases from the $\sim 15$
day separation of the 2 data sets obtained for this object.}
\label{fig:gnpdmsmall}
\end{figure}

\begin{figure}
\caption{The phased BVRI data from the UT April $20-25$ and May
$11-13$, 2001 observations of 2000 GN$_{171}$.  The period has been
phased to 8.329 hours which is the best fit double-peaked period. The
May data have been corrected for geometry and phase angle differences
relative to the April data (Table~1).  The BVI data have been shifted
by the amount indicated on the graph in order to correspond to the R
data.  No color variation is seen within our uncertainties.  A Fourier
fit shows the two pronounced maximum and minimum.}
\label{fig:lightgnbvri}
\end{figure}

\begin{figure}
\caption{The colors of 2000 GN$_{171}$ plotted against rotational phase.}
\label{fig:gncolors}
\end{figure}

\begin{figure}
\caption{Phase dispersion minimization (PDM) plot for (33128) 1998
BU$_{48}$ from the November 2001 data.  Best fits from this plot are
the 4.9 and 6.3 hour single-peaked fits and the 9.8 and 12.6 hour
double-peaked fits.  }
\label{fig:bupdm}
\end{figure}

\begin{figure}
\caption{BVRI phased data from the UT November $14-19$ observations of
(33128) 1998 BU$_{48}$.  The period has been phased to 6.29 hours
which is one of the best fit single-peaked periods for (33128) 1998
BU$_{48}$, the other being around 4.9 hours.}
\label{fig:buphase}
\end{figure}

\begin{figure}
\caption{Phase dispersion minimization (PDM) plot for 1999 KR$_{16}$
using all the R-band data from February, April and May 2001. Best fits from this plot
are near the 5.9 hour single-peak period and the 11.8 hour
double-peaked period.  Both are flanked by aliases of the 24 hr and 
$\sim 15$ and $\sim 60$ day sampling periodicities.}
\label{fig:krpdm}
\end{figure}

\begin{figure}
\caption{A closer view of the phase dispersion minimization (PDM) plot
for 1999 KR$_{16}$ around the best fit single-peaked periods near 5.9
hours.  }
\label{fig:krpdmsmall}
\end{figure}

\begin{figure}
\caption{The phased BVRI data from the UT April $24-25$ and May
$11-13$, 2001 observations of 1999 KR$_{16}$.  The period has been
phased to 5.840 hours which is one of the best fit single-peaked
period for 1999 KR$_{16}$, the other being at 5.929 hours.}
\label{fig:lightkr1}
\end{figure}

\begin{figure}
\caption{The null lightcurves of KBOs found to have no significant
variation: a) 2001 FZ$_{173}$ b) 2001 CZ$_{31}$ c) (38628) 2000
EB$_{173}$ d) (26375) 1999 DE$_{9}$ e) 1998 HK$_{151}$ f) (33340) 1998
VG$_{44}$ g) (19521) Chaos 1998 WH$_{24}$ h) 1997 CS$_{29}$ i) (26181)
1996 GQ$_{21}$.}
\label{fig:lightall}
\end{figure}

\begin{figure}
\caption{Rotational variability and periods of all the asteroids with
diameters $>200$ km and of Kuiper Belt objects in our sample.  Objects
in the upper and upper left portions of the graph are possibly
rotationally deformed rubble piles.  The asteroid amplitudes which
were taken from pole orientations of 90 degrees have been corrected to
a mean pole orientation at 60 degrees to better compare them with the
KBOs of unknown orientation.  KBOs with amplitudes $\leq$ 0.1
magnitudes and periods $\geq$ 12 hours are subject to observational
bias against detection.}
\label{fig:ampper} 
\end{figure}

\begin{figure}
\caption{Monte Carlo simulations using a constant axis ratio for all
KBOs.  Error bars for the KBO points are based on a Poisson
distribution.}
\label{fig:simulation1}
\end{figure}

\begin{figure}
\caption{Monte Carlo simulations using Gaussians centered on the axis
ratio of 1:1.5 with different standard deviations
(Equation~\ref{eq:simmodel3}).  Error bars for the KBO points are
based on a Poisson distribution.}
\label{fig:simulation2a}
\end{figure}

\newpage
\clearpage

\begin{figure}
\caption{Monte Carlo simulations using power laws of different slopes
(Equation~\ref{eq:simmodel4}).  Error bars for the KBO points are
based on a Poisson distribution.}
\label{fig:simulation2b}
\end{figure}

\begin{figure}
\caption{Monte Carlo simulations using all large asteroids ($D>200$
km) and a collisional distribution from Catullo et al. (1984).  Error
bars for the KBO points are based on a Poisson distribution.}
\label{fig:simulation3}
\end{figure}

\begin{figure}
\caption{Sizes and densities of icy bodies.  A trend is observed in
which the larger the object the higher the density.  The solid line is
over plotted to show the expected bulk density of a pure water ice
sphere with size (Lupo and Lewis 1979). Other lines indicate how the
density would behave with added porosity and rock.  Data points for
satellite densities are from the JPL Solar System Dynamics web page.}
\label{fig:iceden} 
\end{figure}

\begin{figure}
\caption{Size and densities of possible rotationally deformed KBOs and
main belt asteroids.  The asteroids have lower densities than expected
for solid rock, but are still denser than the KBOs. }
\label{fig:iceast} 
\end{figure}

\begin{figure}
\caption{Phase functions for Kuiper Belt objects observed at several
phase angles.  The best linear fit gives a phase coefficient of $\beta
(\alpha < 2\arcdeg ) = 0.15$ magnitudes per degree.  Objects with more
than two data points show evidence of the nonlinear opposition surge.}
\label{fig:phaseall} 
\end{figure}

\begin{figure}
\caption{Phase functions of all 7 KBOs observed at multiple phase
angles.  The reduced magnitudes have been normalized to show all
objects relative slopes.  Over plotted are fits of the slope parameter
$G=0.05$, $0.15$ (C-type), and $0.25$ (S-type).  The best fit slope
parameters of all KBOs are below $G=0.05$ which is consistent
with scattering from low albedo surfaces.}
\label{fig:phaseallG}
\end{figure}

\begin{figure}
\caption{Comparison of phase functions for the typical KBO 1999
KR$_{16}$ and Pluto.  The Solid line is the best fit Bowell et
al. $HG$ phase function for 1999 KR16 with $G=-0.08$.  Data points for
Pluto are from Tholen \& Tedesco (1994) and are offset in the vertical
direction from -1.0.  Pluto has a best fit $G=0.88$ shown with the
dashed line. }
\label{fig:phasekr} 
\end{figure}


\begin{references}

\reference{Asp98} Asphaug, E, Ostro, S., Hudson, R., Scheeres, D., \& Benz, W.  1998, Nature, 393, 437

\reference{Bel00} Belskaya, I. \& Shevchenko, V.  2000, Icarus, 147, 94

\reference{Bin89} Binzel, R., Farinella, P., Zappala V., \& Cellino, A.  1989,  in Asteroids II, ed. R. Binzel, T. Gehrels, and M. Matthews (Tucson: Univ. of Arizona Press), 416

\reference{Boe01} Boehnhardt, H., Tozzi G., Birkle, K. et al. 2001, AA, 378, 653

\reference{Bow89} Bowell, E., Hapke, B., Domingue, D., Lumme, K., Peltoniemi, J., \& Harris, A.  1989, in Asteroids II, ed. R. Binzel, T. Gehrels, and M. Matthews (Tucson: Univ. of Arizona Press), 524

\reference{Bro99} Brown, R., Cruikshank, D., \& Pendleton, Y. 1999, ApJ Lett, 519, L101

\reference{Bur73} Burns, J., \& Safronov, V. 1973, MNRAS, 165, 403

\reference{Cap84} Capaccioni, F., Cerroni, P., Coradini, M., Farinella, P., Flamini, E., Martelli, G., Paolicchi, P., Smith, P., \& Zappala, V. 1984, Nature, 308, 832

\reference{Cat84} Catullo, V., Zappala, V., Farinella, P., \& Paolicchi, P. 1984, AA, 138, 464

\reference{Cha69} Chandrasekhar, S. 1987, Ellipsoidal Figures of Equilibrium.  Dover, New York.

\reference{Col99} Collander-Brown, S., Fitzsimmons, A., Fletcher, E., Irwin, M., \& Williams, I. 1999, MNRAS, 308, 588

\reference{Dav97} Davies, J., McBride, N. \& Green, S. 1997, Icarus, 125, 61

\reference{Dea95} De Angelis, G. 1995, PSS, 43, 649

\reference{Deg79} Degewij, J., Tedesco, E., \& Zellner, B. 1979, Icarus, 40, 364

\reference{Dru91} Drummond, J., Weidenschilling, S., Chapman, C., \& Davis, D.  1991, Icarus, 89, 44

\reference{Dur00} Durda, D. \& Stern, A. 2000, Icarus, 145, 220

\reference{Far01} Farnham, T. 2001, IAU Circular 7583 (February 16)

\reference{Far81} Farinella, P., Paolicchi, P., Tedesco, E., \& Zappala, V. 1981, Icarus, 46, 114

\reference{Far82} Farinella, P., Paolicchi, P., \& Zappala, V. 1982, Icarus, 52, 409

\reference{Far96} Farinella, P. \& Davis, D. 1996, Sci, 273, 938

\reference{Fuj78} Fujiwara, A., Kamimoto, G., \& Tsukamoto, A. 1978, Nature, 272, 602

\reference{Geh70} Gehrels, T. 1970, in Surfaces and Interiors of Planets and Satellites, ed. A. Dollfus (London: Academic Press), pg. 355

\reference{Gre97} Green, S., McBride, N., O'Ceallaigh, D. Fitzsimmons, A., Williams, I., \& Irwin, M. 1997, MNRAS, 290, 186

\reference{Hai00} Hainaut, O., Delahodde, C., Boehnhardt, H., Dotto, E., Barucci, M., Meech, K., Bauer, J., West, R., \& Doressoundiram, A. 2000, AA, 356, 1076

\reference{Har94} Harris, A. 1994, Icarus, 107, 209

\reference{Hil97} Hilton, J. 1997, AJ, 114, 402

\reference{Jew93} Jewitt, D. \& Luu, J. 1993, Nature, 362, 730

\reference{Jew96} Jewitt, D., Luu, J., and Chen, J. 1996.  AJ, 112, 1225

\reference{Jew01a} Jewitt, D., Aussel, H., \& Evans, A. 2001, Nature, 411, 446

\reference{Jew01b} Jewitt, D. \& Luu, J. 2001, AJ, 122, 2099

\reference{Jew01c} Jewitt, D. \& Sheppard, S. 2002, AJ, 123, 2110

\reference{Ken99} Kenyon, S. \& Luu, J. 1999, AJ, 118, 1101

\reference{Lan92} Landolt, A. 1992, AJ, 104, 340

\reference{Lar89} Lagerkvist, C., Harris, A., \& Zappala, V. 1989, in Asteroids II, ed. R. Binzel, T. Gehrels, and M. Matthews (Tucson: Univ. of Arizona Press), 1162

\reference{Lei00} Leinhardt, Z., Richardson, D., \& Quinn, T. 2000, Icarus, 146, 133

\reference{Lup79} Lupo, M. \& Lewis, J. 1979, Icarus, 40, 157

\reference{Luu96} Luu, J. \& Jewitt, D. 1996, AJ, 112, 2310

\reference{Luu98} Luu, J. \& Jewitt, D. 1998, AJ, 494, L117

\reference{Mer99} Merline, B., Close, L., Dumas, C., Chapman, C., Roddier, F., Menard, F., Slater, D., Duvert, G., Shelton, C., \& Morgan, T. 1999, Nature, 401, 565

\reference{Ort01} Ortiz, J., Lopez-Moreno, J., Gutierrez, P., \& Baumont, S. 2001, BAAS, 33, 1047

\reference{Rom99} Romanishin, W. \& Tegler, S. 1999, Nature, 398, 129

\reference{Rom01} Romanishin, W., Tegler, S., Rettig, T., Consolmagno, G., \& Botthof, B. 2001, Proc. Nat. Academy Sci., 98, 11863

\reference{Rya99} Ryan, E., Davis, D., \& Giblin, I.  1999, Icarus, 142, 56

\reference{Sca80} Scaltriti, F. \& Zappala, V.  1980, AA, 83, 249

\reference{Sek02} Sekiguchi, T., Boehnhardt, H., Hainaut, O., \& Delahodde, C. 2002, AA, 385, 281

\reference{She02} Sheppard, S. 2002, in preparation

\reference{Ste78} Stellingwerf, R. 1978, ApJ, 224, 953

\reference{Ted89} Tedesco, E. 1989, in Asteroids II, ed. R. Binzel,
T. Gehrels, and M. Matthews (Tucson: Univ. of Arizona Press), pg. 1090

\reference{Teg00} Tegler, S. \& Romanishin, W. 2000, Nature, 407, 979

\reference{Tho94} Tholen, D. \& Tedesco E. 1994, Icarus, 108, 200

\reference{Via00} Viateau, B. 2000, AA, 354, 725

\reference{Yeo97} Yeomans, D., Barriot, J., Dunham, D. et al. 1997,
Science, 278, 2106

\end{references}
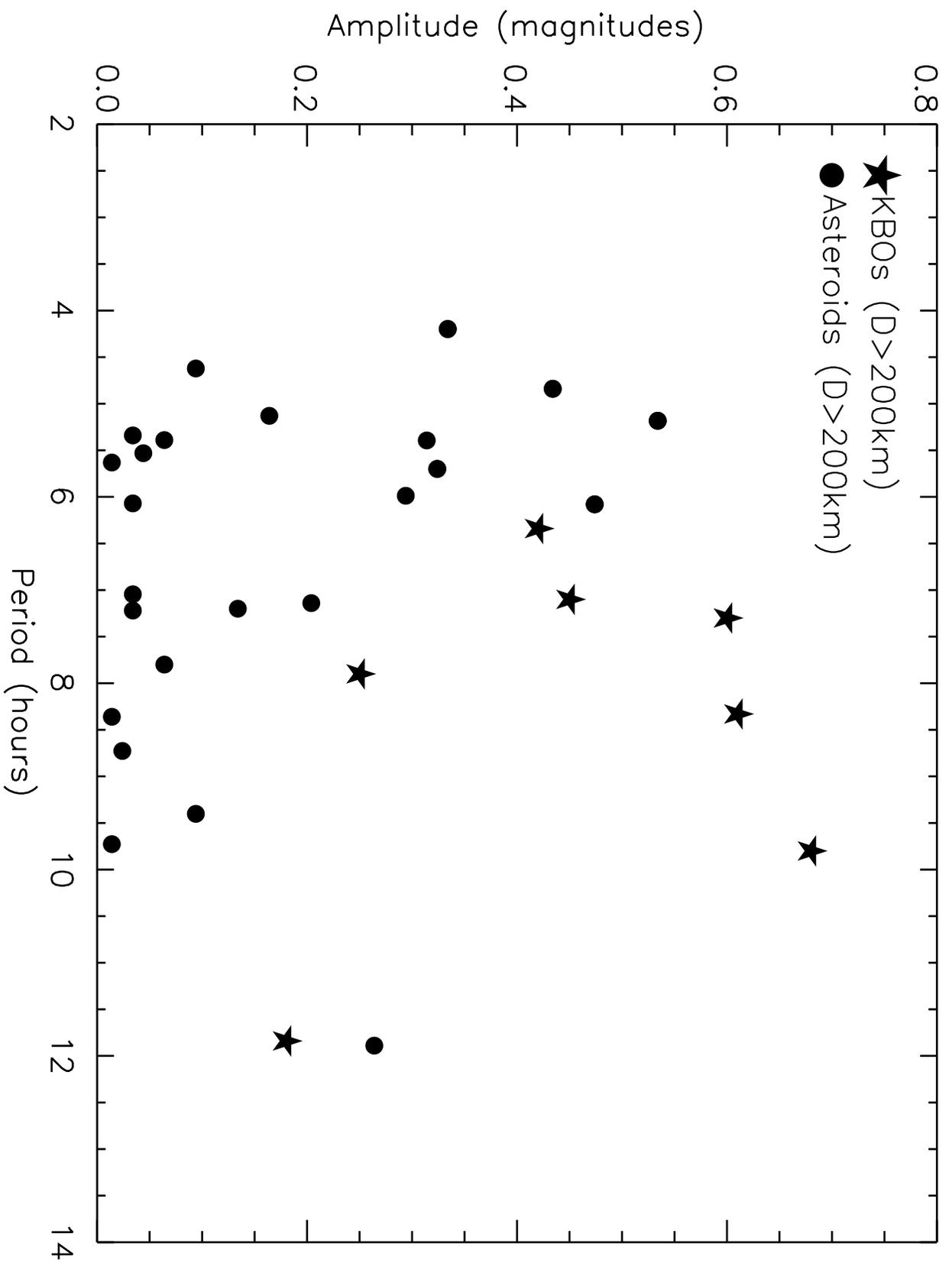
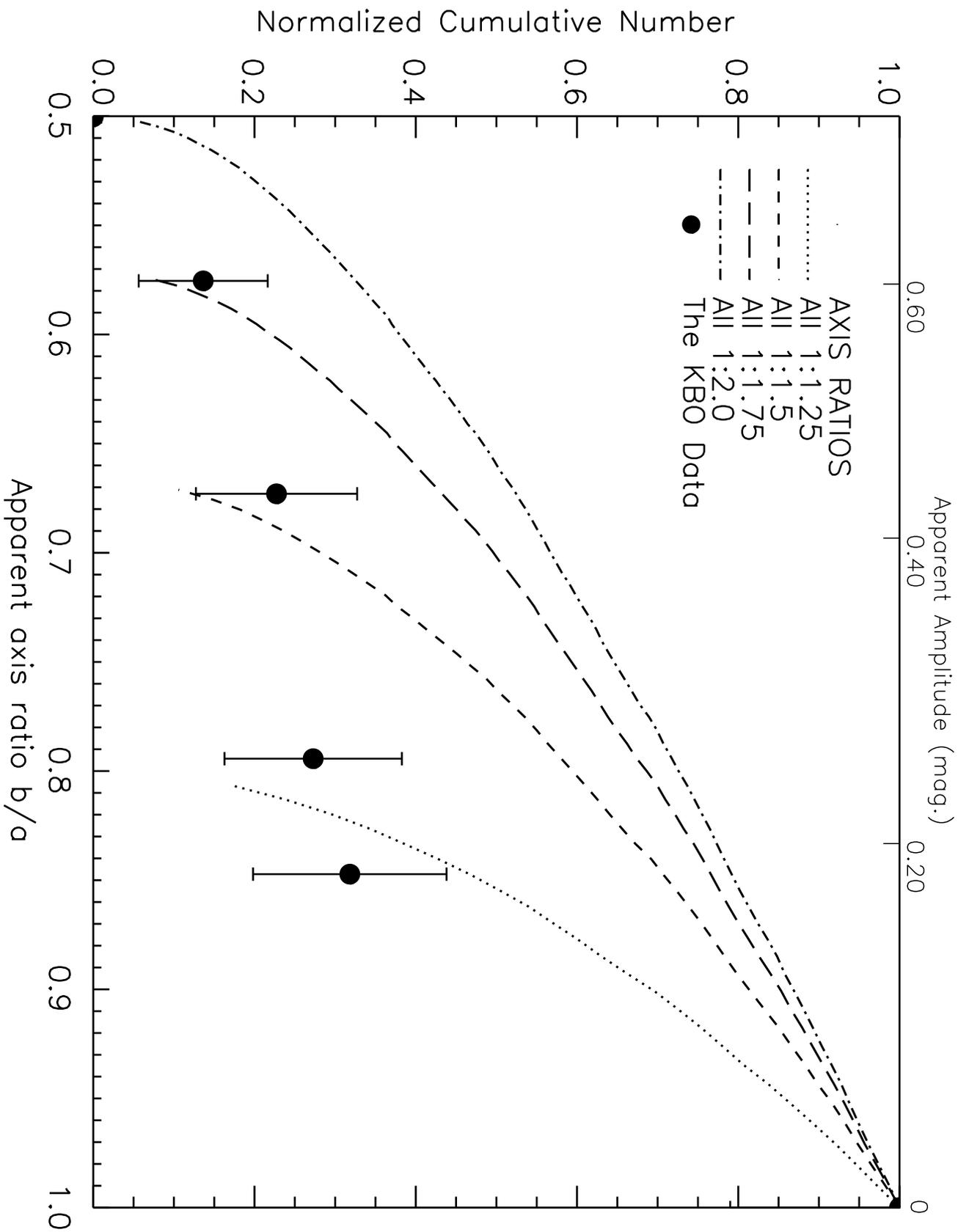
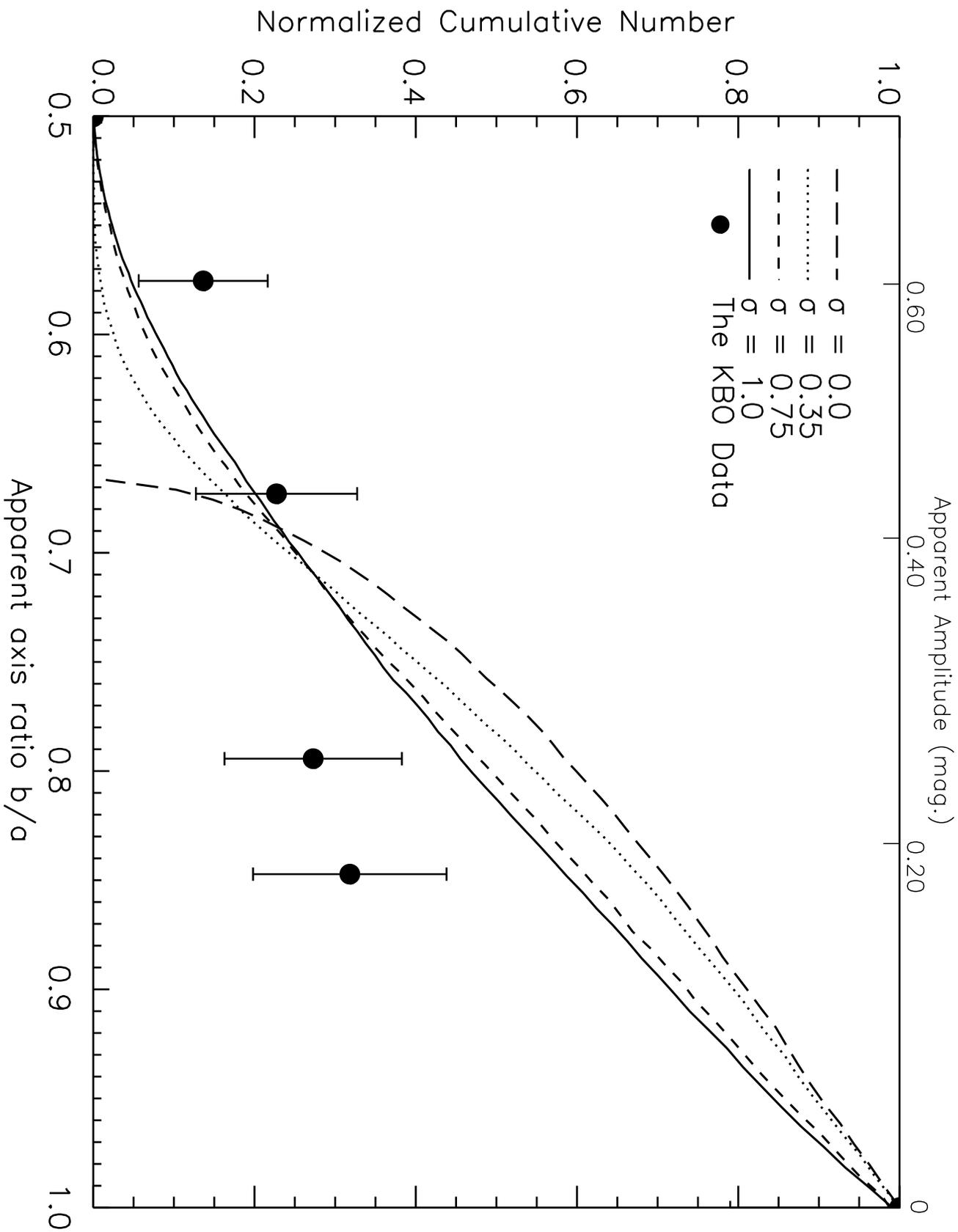
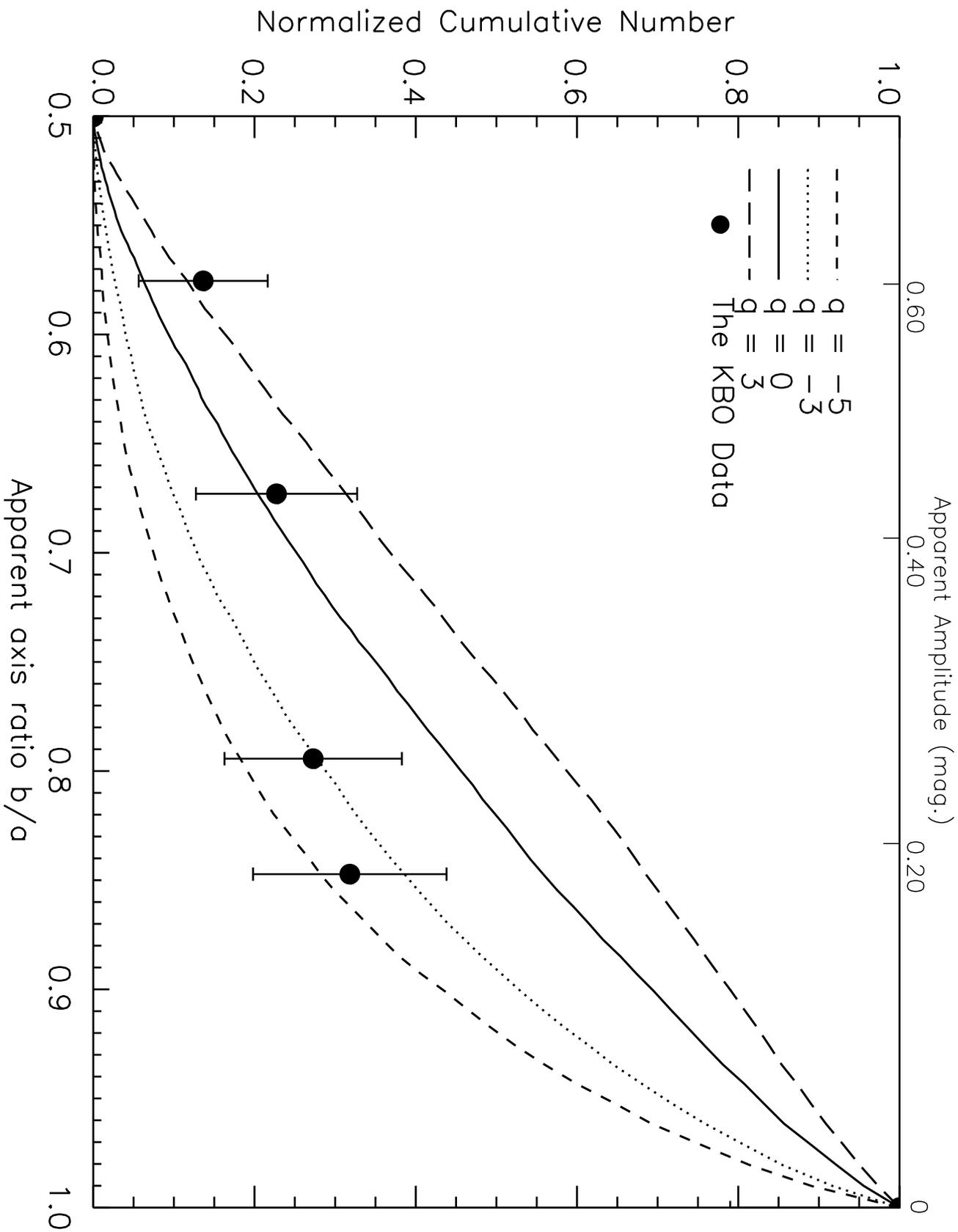
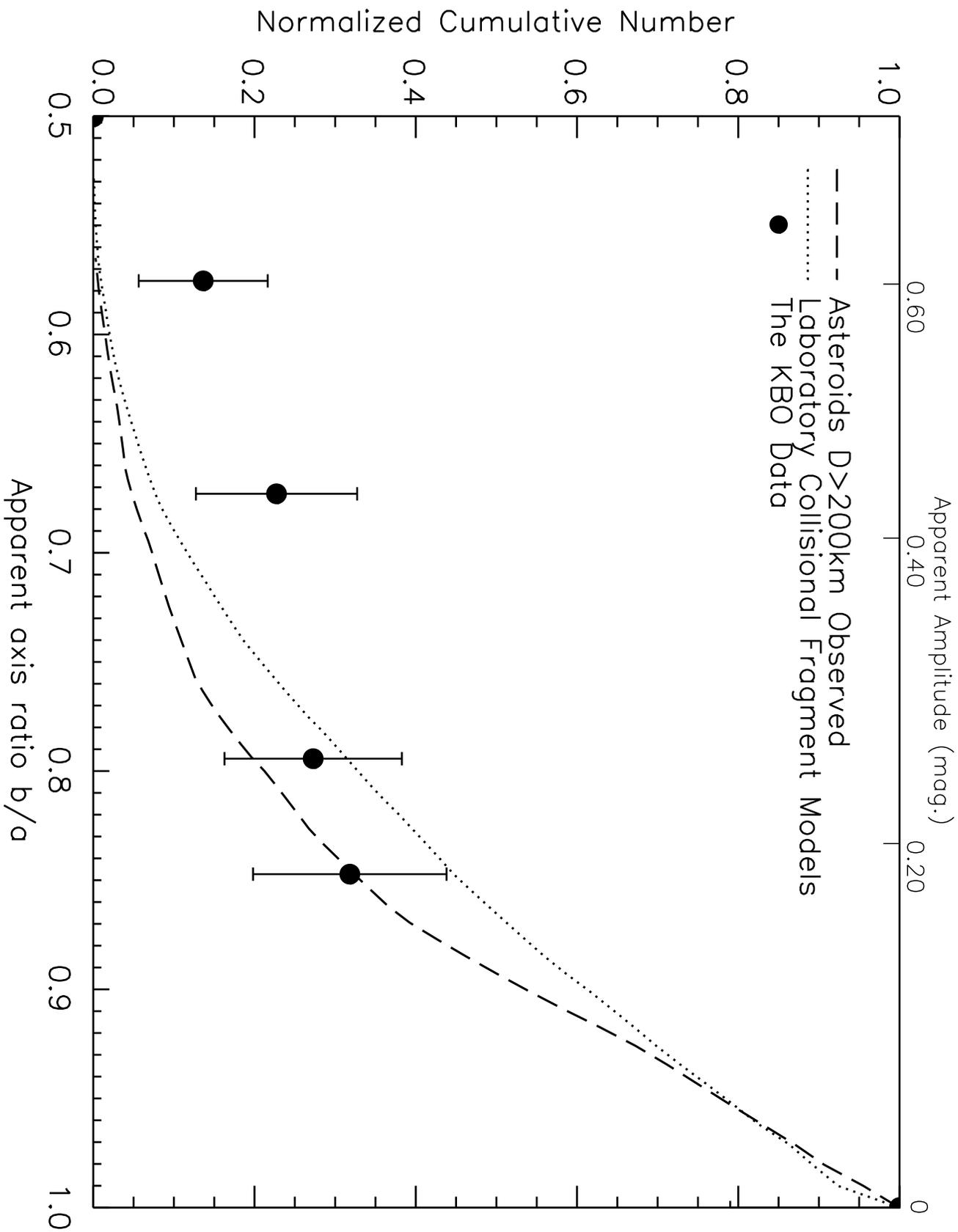
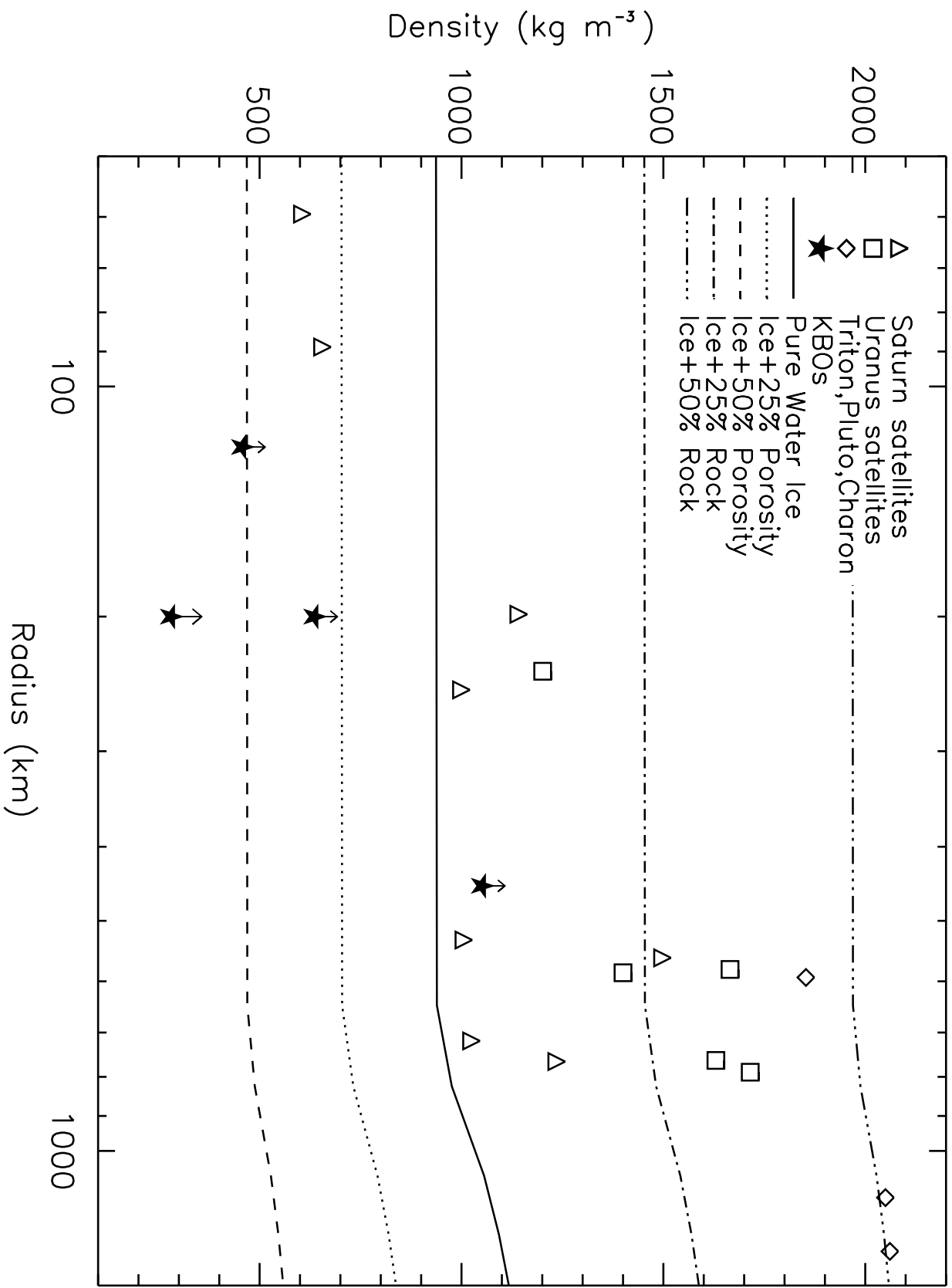
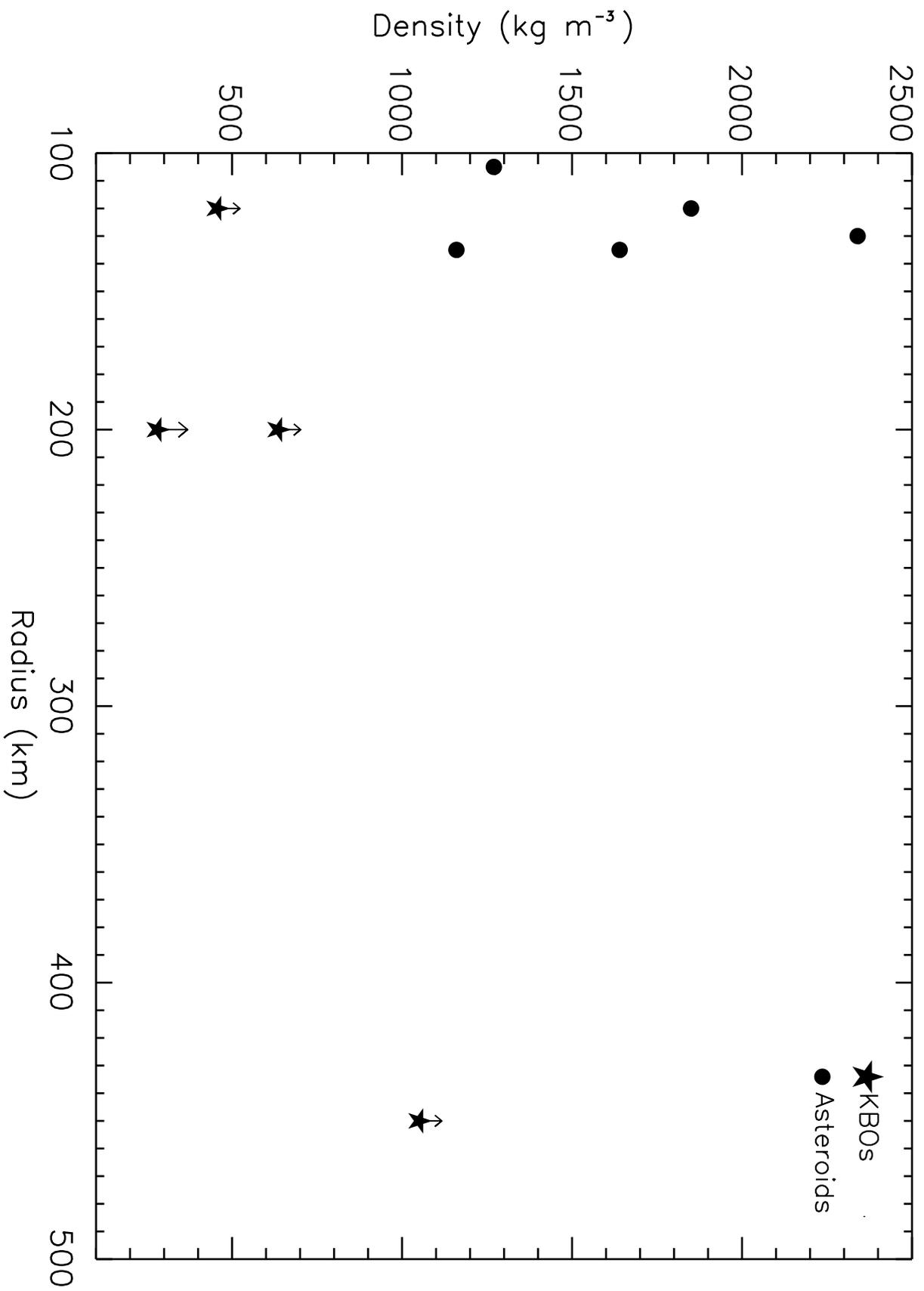
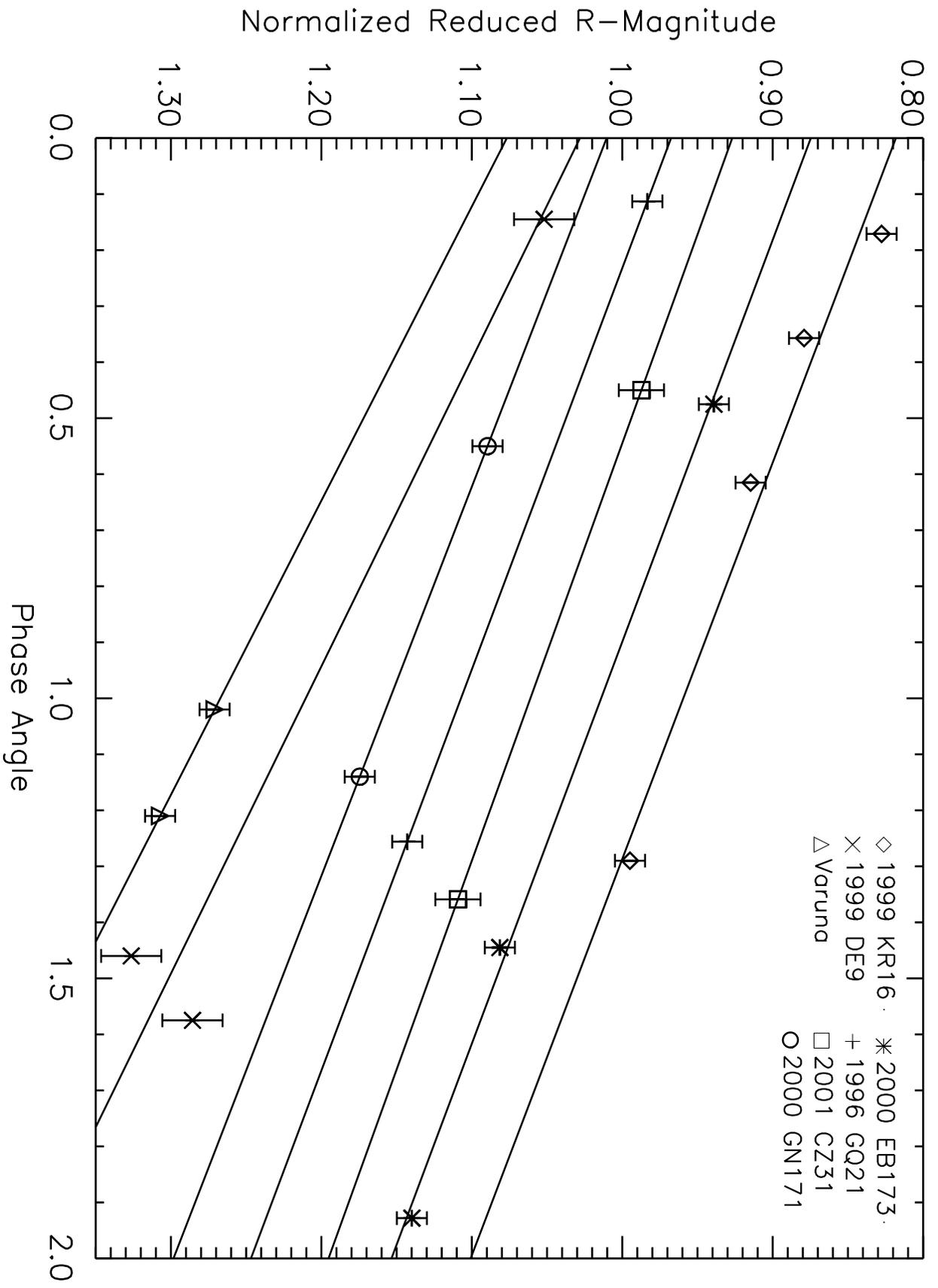
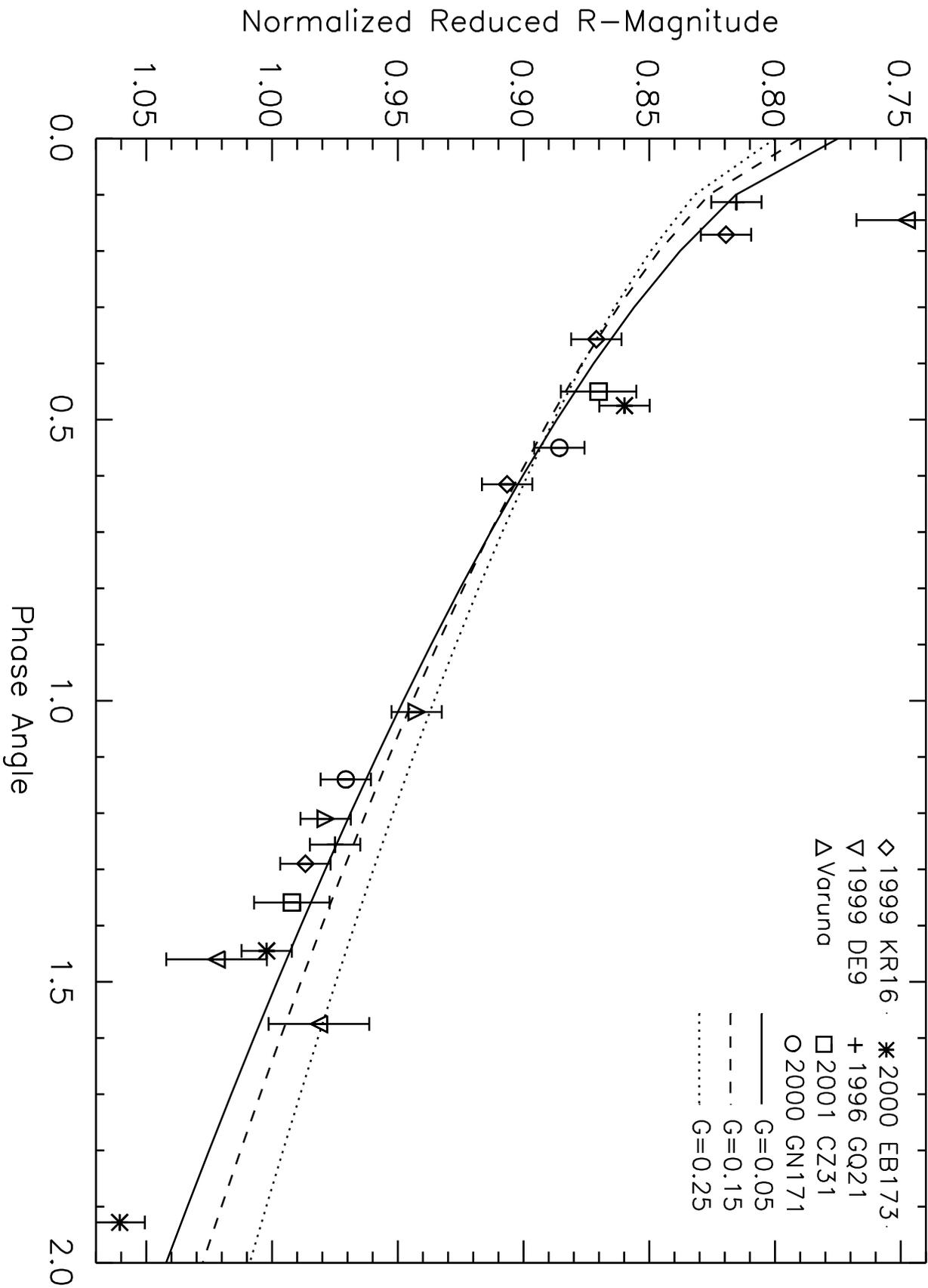
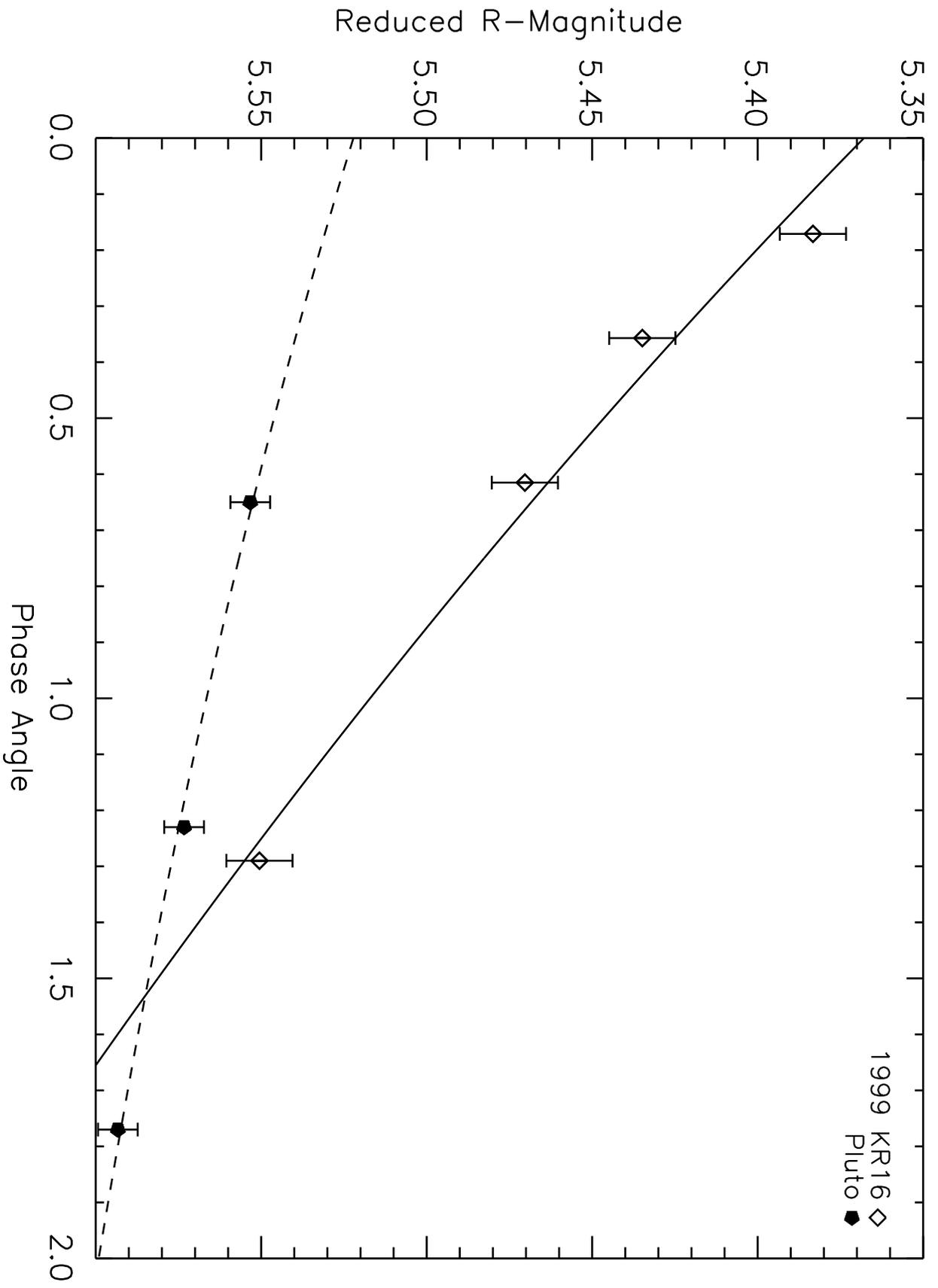
\end{document}